\documentclass[aps,prl,twocolumn,preprintnumbers,showpacs]{revtex4}        
\usepackage{graphicx}
\usepackage{amsfonts}
\usepackage{amsmath}
\usepackage{amssymb}
\usepackage{amsfonts}
\newcommand{\p}{\partial}

\begin{document}
\title{Phase space reduction of the one-dimensional Fokker-Planck
(Kramers) equation}

\author{Pavol Kalinay$^{1}$ and Jerome K. Percus$^{2,3}$}

\affiliation{$^{1}$Institute of Physics, Slovak Academy of
Sciences, D\'ubravska cesta 9, 84511, Bratislava, Slovakia\\
$^{2}$Courant Institute of Mathematical Sciences, New York
University, New York, NY, 10012 \\ $^{3}$Department of Physics, New
York University, 4 Washington Place, New York, New York 10003}

\date{}

\begin{abstract}
A pointlike particle of finite mass $m$, moving in a one-dimensional viscous
environment and biased by a spatially dependent force, is considered.
We present a rigorous mapping of the Fokker-Planck equation, which
determines evolution of the particle density in phase space, onto
the spatial coordinate $x$. The result is the Smoluchowski equation,
valid in the overdamped limit, $m\rightarrow 0$, with a series of
corrections expanded in powers of $m$. They are determined unambiguously
within the recurrence mapping procedure. The method and the results are
interpreted on the simplest model with no field and on the damped harmonic
oscillator.
\end{abstract}


\maketitle

\renewcommand{\theequation}{1.\arabic{equation}}
\setcounter{equation}{0}

\section{I. Introduction}

The Brownian motion of a particle in a confined system represents an
essential model used in description of stochastic transport through
quasi one-dimensional (1D) systems, e.g. channels in nanomaterials,
pores or fibers in biological structures. In a 1D system, the
trajectory $x(t)$ of a Brownian particle is described by the
Langevin equation
\begin{equation}\label{1.1}
m\ddot x+\gamma\dot x+\partial_xU(x)=f(t)\ .
\end{equation}
Here, $m$ denotes mass of the particle, $U(x)$ represents the driving
potential, $\gamma$ is an effective friction coefficient and $f(t)$ is the
stochastic force, satisfying the usual conditions on averaged values,
$\left< f(t)\right>=0$ and $\left<f(t)f(t')\right>=2\gamma k_BT
\delta(t-t')$; $T$ is the temperature and $k_B$ the Boltzmann
constant. The corresponding phase space density $\rho(x,v,t)$ of the
particle satisfies the Fokker-Planck (FP), or Kramers (kinetic) equation,
\begin{eqnarray}\label{1.2}
\left(\partial_t+v\partial_x-\frac{U'(x)}{m}\partial_v\right)
\rho(x,v,t)&& \cr && \hskip-1.5in =\frac{\gamma}{\beta m^2}\partial_v
e^{-\beta mv^2/2}\partial_v e^{\beta mv^2/2}\rho(x,v,t),\hskip0.3in
\end{eqnarray}
where $x$ is the spatial coordinate, $v$ denotes its velocity and
$\beta=1/k_BT$ is the inverse temperature.

Solutions of the Langevin equation, as well as the corresponding kinetic
equation, are studied over almost a century \cite{Chand,Risken,HangR}.
Still, motion of a particle in confined geometries represents usually a
complicated problem, requiring next reductions of the used description.
Due to simplicity, and often for relevance, mainly the overdamped limit
is studied. The mass dependent term in Eq. (\ref{1.1}), $m\ddot x$, is
then considered negligible and the particle's spatial density $p(x,t)$
is governed by the Smoluchowski equation,
\begin{equation}\label{1.3}
\partial_tp(x,t)=D_0\partial_xe^{-\beta U(x)}\partial_x
e^{\beta U(x)}p(x,t)\ ,
\end{equation}
containing no information about the mass of the particle; $D_0$ denotes the
diffusion constant. Instead of the full phase space, one works with only
the spatial coordinate $x$. Of course, the question is the price for this
simplification, as well as possibility of also including properly inertia
of the massive particles in the reduced (real space only) description of
the Brownian dynamics. Recent studies showed its importance in understanding
rectification of the transport in ratchets \cite{MarchSN}, or its influence
on the effective diffusion coefficient \cite{Ber,GhH,MarS} in a narrow channel.
For our purpose to demonstrate the phase space reduction, we will deal here
with only the 1D FP (Kramers) equation (\ref{1.2}).

The eq. (\ref{1.2}) is reducible to the Smoluchowski eq. (\ref{1.3}) in the
limit $m\rightarrow 0$. The reduction procedure \cite{Smol,Kram,Gard} is
based on an instant thermalization of the particle's velocity after any
move in the $x$ direction in the case of an infinitely small mass. The
situation resembles derivation of the Fick-Jacobs (FJ) equation \cite{FJ,20},
\begin{equation}\label{1.4}
\partial_tp(x,t)=\partial_x A(x)\partial_x\frac{p(x,t)}{A(x)},
\end{equation}
appearing as the result of the dimensional reduction of the diffusion
equation in a 2D channel with varying cross section $A(x)$, onto the
longitudinal coordinate $x$, if equilibration in the transverse
($y$) direction is instant. The function $p(x,t)$ denotes the
linear (1D) density of the particle. Of course, as in reduction of
diffusion to the FJ equation, that of (\ref{1.2}) to (\ref{1.3}) as
$m\rightarrow 0$ is a singular limit, and must be handled with care, but
from the viewpoint of the reduction of (\ref{1.1}), no such caveat is needed. 

Recently, an exact mapping procedure was proposed \cite{map,exact},
enabling us to also derive the corrections to the FJ equation (\ref{1.4}),
which are necessary, if the transverse equilibration is not instant.
The key was to introduce anisotropy of the diffusion constant in the
diffusion equation,
\begin{equation}\label{1.5}
\partial_t\rho(x,y,t)=\left(\partial_x^2+\frac{1}{\epsilon}
\partial_y^2\right)\rho(x,y,t),
\end{equation}
governing the 2D (spatial) density $\rho(x,y,t)$. For $\epsilon\rightarrow
0$, the infinitely fast transverse diffusion immediately flattens the
$y$ profile of $\rho(x,y,t)\rightarrow\rho(x,t)=p(x,t)/A(x)$. Then
integration of Eq. (\ref{1.5}) over the cross section, together with
the reflecting boundary conditions satisfied at the hard walls, results
in the FJ eq. (\ref{1.4}). In the case of a slower transverse diffusion,
the mapping procedure generates a series of corrections to the FJ equation
controlled by $\epsilon>0$, developing the $y$ profile of $\rho(x,y,t)$, 
which is already curved depending on the flux and geometry of the channel.

The procedure was extended to mapping of diffusion in an external field
$U(x,y)$, e.g. for diffusion in a channel with soft walls \cite{soft},
where the particle is kept near the $x$ axis by the parabolic potential,
$\beta U(x,y)=\alpha(x)y^2$; $\alpha(x)$ represents the varying
stiffness of the walls. The equation to be mapped onto the $x$
coordinate is the 2D Smoluchowski equation,
\begin{eqnarray}\label{1.6}
\left(\partial_t-\partial_xe^{-\alpha(x)y^2}\partial_x
e^{\alpha(x)y^2}\right)\rho(x,y,t)=\hskip0.8in\cr
=\frac{1}{\epsilon}\partial_ye^{-\alpha(x)y^2}\partial_y
e^{\alpha(x)y^2}\rho(x,y,t).\hskip0.2in
\end{eqnarray}
Integrating over $y$ and applying the mapping scheme gives the mapped
1D equation in an extended Smoluchowski form,
\begin{equation}\label{1.7}
\partial_tp(x,t)=\partial_xe^{-V(x)}\left[1+
\epsilon\hat Z(x,\partial_x)\right]\partial_xe^{V(x)}p(x,t),
\end{equation}
governing the 1D density $p(x,t)$, where $V(x)$ stands for an
effective potential and $\hat Z$ is the correction operator sought as
an expansion in the small parameter $\epsilon$. The potential $V(x)$
and the coefficients of $\hat Z$ depend on $\alpha(x)$ and both are
fixed unambiguously within the recurrence mapping procedure.

The central idea of this paper is a conjecture that the FP eq.
(\ref{1.2}) can be reduced to an extended Smoluchowski-like 1D form,
governing the spatial density $p(x,t)$, after integration over $v$ and
applying a similar mapping scheme. The velocity $v$ thus represents a
"transverse" coordinate instead of $y$, with the mass $m$ playing the
role of the small parameter $\epsilon$. Then the Smoluchowski equation
(\ref{1.3}) should be obtained in the limit $m\rightarrow 0$ \cite{Smol,Kram}.
Performing the recurrence procedure, a series of corrections to this
equation in powers of $m$ would be derived. Then the final mapped equation
would also respect inertia of the Brownian particles, although working only
in real space.

Use of the mapping procedure developed for diffusion \cite{map,
exact,soft,eff} is not straightforward; the left-hand side operator of
the FP equation (\ref{1.2}) has a different structure than that of
the diffusion (\ref{1.5}) or Smoluchowski equation (\ref{1.6}). Still,
there is a way to apply the general scheme of the mapping in this case
and to perform reduction of the phase space onto the real space in
the way described above. Presentation of this algorithm is the primary
aim of this study.

The result of the mapping of the FP equation (\ref{1.2}) is again an
equation of the form (\ref{1.7}) with $V(x)=\beta U(x)$ and $\epsilon$
replaced by $m$. In the limit of stationary flow, this equation can
be simplified by subsequent reduction of the operator $1-m\hat Z$ to a
function $D(x)$, a spatially dependent effective diffusion coefficient
\cite{20,RR},
\begin{equation}\label{1.8}
\partial_tp(x,t)=\partial_xe^{-\beta U(x)}D(x)\partial_x
e^{\beta U(x)}p(x,t)\ .
\end{equation}
The leading term of $D(x)$ is proportional to $mU''(x)$ and the
whole series of corrections to the Smoluchowski equation can be summed
up, giving
\begin{equation}\label{1.9}
D(x)=D_0\frac{1-\sqrt{1-4mU''(x)/\gamma^2}}{2mU''(x)/\gamma^2},
\end{equation}
with $D_0=1/\gamma\beta$, if the higher derivatives of $U(x)$ are neglected.

In the following section, we analyze how to apply the mapping scheme
for reduction of the phase space in the FP equation (\ref{1.2}). In
Sect. III, our considerations are verified on an exactly solvable model,
the FP equation with no field, $U(x)=0$. Analysis of this example helps
us to construct the recurrence scheme for calculation of the series of
corrections to the zeroth order approximation, the Smoluchowski eq.
(\ref{1.3}), in the small parameter $m$. Finally, the complete mapping
procedure for an arbitrary potential $U(x)$ is presented in Sect. IV.
The mapped equation of type (\ref{1.7}), as well as the formula (\ref{1.9})
is derived, and checked on the damped linear harmonic oscillator.

\renewcommand{\theequation}{2.\arabic{equation}}
\setcounter{equation}{0}

\section{II. Preliminary considerations}

The key points of the mapping procedure, as formulated for diffusion
\cite{map,exact}, are recalled in this Section. Based on physical
considerations, we adjust the general scheme of the mapping for
dimensional reduction of the FP equation (\ref{1.2}) to this situation.

The mapping procedure represents a consistent transition from a fine
grain to a coarse grain description of some evolution process. The
process is described in details by some partial differential equation
(PDE), governing the density of particles $\rho({\bf r},t)$ in the full
space, defined by the coordinates ${\bf r}$. The dimensional reduction
projects this equation onto another PDE, which governs the density
$p(x,t)$ in the reduced space of the coordinate $x$.
The coordinate $x$ is one of the coordinates of the full space,
${\bf r}=(x,{\bf y})$, and the mapping accomplishes integration over the
transverse coordinates ${\bf y}$. In the case of the FP equation
(\ref{1.2}), phase space $(x,v)$ represents the full space and
the dimensional reduction integrates over the "transverse" coordinate,
the velocity $v$. Hence we have the defining relation between the
densities $\rho$ and $p$:
\begin{equation}\label{2.1}
p(x,t)=\int_{-\infty}^{\infty}\rho(x,v,t)dv\ .
\end{equation}
The phase-space density $\rho(x,v,t)$ is expected to be near the thermodynamic
equilibrium and so approximately proportional to $\exp(-\beta mv^2/2)$,
which provides convergence of the integral in Eq. (\ref{2.1}).

Then the first step of the mapping is also integration of Eq. (\ref{1.2})
over $v$. If completed, we get
\begin{equation}\label{2.2}
\int_{-\infty}^{\infty}\left(\partial_t+v\partial_x\right)\rho(x,v,t)dv
=0;
\end{equation}
the other terms are zero due to $\rho(x,v,t)\rightarrow 0$ in the limit
$v\rightarrow\pm\infty$. This equation represents nothing but mass
conservation in the reduced space,
\begin{equation}\label{2.3}
\partial_tp(x,t)+\partial_xj(x,t)=0,
\end{equation}
where the 1D flux $j$ is defined by the relation
\begin{equation}\label{2.4}
j(x,t)=\int_{-\infty}^{\infty}v\rho(x,v,t)dv\ .
\end{equation}

In contrast to diffusion, where $j=-D_0\partial_xp$ is fixed, here
the flux $j$ is a function formally independent of the density $p$,
so we also need the evolution equation for this quantity. After
integration of Eq. (\ref{1.2}) multiplied by $v$, we obtain
\begin{eqnarray}\label{2.5}
\partial_tj(x,t)+\partial_x\int_{-\infty}^{\infty}v^2\rho(x,v,t)dv
+\frac{U'(x)}{m}p(x,t)=\ \ \ \cr =-\frac{\gamma}{\beta m^2}
\int_{-\infty}^{\infty}e^{-\beta mv^2/2}\partial_ve^{\beta mv^2/2}
\rho(x,v,t)dv\ \ 
\end{eqnarray}
[after some algebra and applying the definitions (\ref{2.1}) and
(\ref{2.4}). This step recalls the Grad's method of moments
\cite{Grad,Rein}. For the 1D Kramers equation (\ref{1.2}), taking only
a couple of the zeroth ($p$) and the first ($j$) order moment of the
phase space density $\rho$ is satisfactory for generating the
selfconsistent system of the mapped (real space) equations (\ref{2.3})
and (\ref{2.5}).]

The next key point of the mapping algorithm is that of expressing the
full-space density $\rho(x,v,t)$ using the 1D density $p(x,t)$ and
also the flux $j(x,t)$ in this case. This relation enables us to
complete integrations in Eq. (\ref{2.5}) and get the evolution
equation for $j(x,t)$, together with the 1D mass conservation, in
closed form. The initial task is to find the zeroth order approximation,
valid in the limit $m\rightarrow 0$. Our first proposal for such a relation
between $\rho(x,v,t)$ and the reduced space quantities $p$ and $j$
is based on the following physical construction:

For an infinitely small mass of the particle, the stochastic force
thermalizes its velocity $v$ almost immediately after any move
along the spatial coordinate $x$. Similar to the transverse
equilibration of the 2D density of a particle diffusing in a narrow
channel with biasing transverse force \cite{soft,Bur1,Mart,gravi},
one could try the formula with separated Boltzmann factor in the
fast relaxing "direction" $v$,
$\rho(x,v,t)\simeq\sqrt{\beta m/2\pi}p(x,t)\exp(-\beta mv^2/2)$.
It is easy to check that it does not work here; the flux $j$ becomes
zero according to Eq. (\ref{2.4}). To prevent this failure, let us
suppose that the distribution in $v$ is shifted by the local mean
(macroscopic) velocity $v_0$, depending on the local flux,
$j(x,t)=v_0(x,t)p(x,t)$. Then we have
\begin{eqnarray}\nonumber
\rho(x,v,t)&\simeq& \sqrt{\frac{\beta m}{2\pi}}
e^{-\beta m(v-v_0)^2/2}p(x,t)\cr
&\simeq&\sqrt{\frac{\beta m}{2\pi}}\left[1+\beta mvv_0+ ...\right]
e^{-\beta mv^2/2}p(x,t).
\end{eqnarray}
Retaining only these two terms in the square brackets and replacing
$v_0p$ by the flux $j$, one gets
\begin{equation}\label{2.6}
\rho_0(x,v,t)=\sqrt{\frac{\beta m}{2\pi}}e^{-\beta mv^2/2}
\left[p(x,t)+\beta mvj(x,t)\right], 
\end{equation}
which will be taken as the sought zeroth order relation between $\rho$
and the reduced space quantities, $p$ and $j$.

This heuristic formula will be verified later by the exact mapping
algorithm. Still, one can check immediately that the relation (\ref{2.6})
represents correctly a kind of backward mapping of the 1D (spatial)
functions $p$ and $j$ onto the phase space densities $\rho$; if substituted
for $\rho(x,v,t)$ in the defining relations (\ref{2.1}) and (\ref{2.4}),
we obtain identities. Applying Eq. (\ref{2.6}) to Eq. (\ref{2.5}), the
integrals over $v$ can be completed and the result,
\begin{equation}\label{2.7}
\partial_tj(x,t)+\partial_x\frac{p(x,t)}{\beta m}+\frac{U'(x)}{m}
p(x,t)=-\frac{\gamma}{m}j(x,t),
\end{equation}
together with the mass conservation, Eq. (\ref{2.3}), forms a closed
couple of PDE, governing the mapped quantities $p$ and $j$.

In the limit $m\rightarrow 0$, the first term in Eq. (\ref{2.7}),
$\partial_tj$, becomes negligible and the equation expresses the
zeroth order relation between the flux $j$ and the density $p$,
\begin{equation}\label{2.8}
j(x,t)=-\frac{1}{\beta\gamma}e^{-\beta U(x)}\partial_x
e^{\beta U(x)}p(x,t)\ .
\end{equation}
If combined with the mass conservation, Eq. (\ref{2.3}), we get the
Smoluchowski equation (\ref{1.3}); $1/\beta\gamma=D_0$ represents the
diffusion constant.

The calculation presented shows that the Smoluchowski equation
(\ref{1.3}) is related to the FP equation (\ref{1.2}) in the same
way as the Fick-Jacobs equation \cite{FJ} to the diffusion equation
valid in a narrow 2D channel. Both mapped equations describe an
asymptotic behavior of the full-space density infinitely rapidly
equilibrating in the transverse direction; the velocity $v$ plays
the role of the transverse coordinate for the FP equation. Our
considerations indicate that the mass of the particle, $m$, becomes
the small parameter, controlling the series of corrections to the
Smoluchowski equation in the case when the transverse
equilibration is not infinitely fast.

Following the scheme of the mapping procedure \cite{map,exact}, the
next point is that of searching for the true relation between the phase
space density $\rho(x,v,t)$ and the 1D quantities $p(x,t)$ and $j(x,t)$,
replacing the heuristic formula (\ref{2.6}), valid for nonzero $m$.
Without losing generality, it can be written in the form
\begin{eqnarray}\label{2.9}
\rho(x,v,t)=\sqrt{\frac{\beta m}{2\pi}}e^{-\beta mv^2/2}
\Big[\hat\omega(x,v)p(x,t)\hskip0.4in\cr
\hskip-0.4in+\beta mv\hat\eta(x,v)j(x,t)\Big].
\end{eqnarray}
If the operators $\hat\omega$ and $\hat\eta$ (with $\partial_x$ implicit)
are expandable in $m$, one can substitute for $\rho(x,v,t)$ in the FP
equation (\ref{1.2}) and fix the coefficients of these operators to satisfy
this equation in each order of $m$, similar to the mapping of diffusion.
Then, using the relation of backward mapping (\ref{2.9}) in Eq.
(\ref{2.5}) gives the expansion of the evolution equation for $j$
and finally, in combination with mass conservation (\ref{2.3}),
the sought series of corrections to the Smoluchowski equation
(\ref{1.3}) in terms of the finite mass $m$.

To verify whether this scheme is viable, we analyze the exactly
solvable case with $U(x)=0$ in the next Section.

\renewcommand{\theequation}{3.\arabic{equation}}
\setcounter{equation}{0}

\section{III. Exactly solvable model}

The exact solution of the FP equation (\ref{1.2}) with no potential,
$U(x)=0$, is presented in this Section. We demonstrate the mapping
on the example of the phase space density $\rho(x,v,t)$ evolving from
the initial density $\rho(x,v,0)$ with thermalized velocity $v$.
The mapped equation, as well as the form of the operators $\hat\omega$
and $\hat\eta$ in Eq. (\ref{2.9}) can be found explicitly in this case.
This analysis will direct us in construction of the recurrence
mapping scheme in Sect. IV.

The case $U(x)=0$ is exactly solvable \cite{Nel}, the Green's
function $G(x,v,t;x',v',t')$ of the FP equation (\ref{1.2}),
\begin{eqnarray}\label{3.1}
&&\hskip-0.2in\left[\partial_t+v\partial_x-\frac{\gamma}{\beta m^2}
\partial_ve^{-\beta mv^2/2}\partial_ve^{\beta mv^2/2}\right]
G(x,v,t;x',v',t')\cr
&&\hskip1in=\delta(x-x')\delta(v-v')\delta(t-t')\ ,
\end{eqnarray}
can be derived explicitly (see Appendix A),
\begin{eqnarray}\label{3.2}
G&=&\frac{\gamma\beta}{4\pi}\frac{\Theta(\tau-\tau')}
{\sqrt{\tau-\tau'-\tanh(\tau-\tau')}\sqrt{1-q^2}}\cr
&&\times\exp\bigg(-\frac{\left[2(\xi-\xi')-(u+u')\tanh(\tau-\tau')
\right]^2}{4[\tau-\tau'-\tanh(\tau-\tau')]}\ \ \cr &&\hskip0.4in
-\frac{q}{1-q^2}\left[q(u^2+u'^2)-2uu'\right]-u^2\bigg)\ ,
\end{eqnarray}
if expressed in the scaled coordinates,
\begin{eqnarray}\label{3.3}
\tau&=&\gamma t/2m,\cr
\xi&=&(\beta m/2)^{3/2}\frac{\gamma x}{\beta m^2},\cr
u&=&\sqrt{\beta m/2}\ v,
\end{eqnarray}
and $q=\exp[-2(\tau-\tau')]$. If the thermalized particle is inserted
at time $t=0$ with a spatial distribution $p_0(x)$,
\begin{equation}\label{3.4}
\rho(x,v,0)=\sqrt{\frac{\beta m}{2\pi}}p_0(x)e^{-\beta mv^2/2},
\end{equation}
evolution of the density $\rho$ is given by the formula
\begin{eqnarray}\label{3.5}
\rho(x,v,t)&=&\sqrt{\frac{\beta m}{2\pi}}\int_{-\infty}^{\infty}dv'
\int dx' G(x,v,t;x',v',0)\ \ \cr &&\hskip1in
\times p_0(x')e^{-\beta mv'^2/2}\cr
&=&\int \frac{p_0(x')dx'}{4\pi D_0\sqrt{Z}}\exp\bigg(
-\frac{[\xi-\xi']^2}{Q}\cr &&\hskip0.2in-\frac{Q}{Z}
\left[u-\frac{(1-e^{-2\tau})}{2Q}(\xi-\xi')\right]^2\bigg);
\end{eqnarray}
$D_0=1/\gamma\beta$, the abbreviations
\begin{eqnarray}\label{3.6}
Q&=&\tau-\frac{1}{2}\left(1-e^{-2\tau}\right),\cr
Z&=&Q-\frac{1}{4}\left(1-e^{-2\tau}\right)^2
\end{eqnarray}
are used and the integration over $x'$ runs over the whole (unspecified)
1D spatial domain.

Then the spatial (1D) density $p$ and the flux $j$ are integrated directly
according to Eqs. (\ref{2.1}) and (\ref{2.4}),
\begin{eqnarray}\label{3.7}
p(x,t)&=&\sqrt{\frac{2}{\beta m}}\int\frac{p_0(x')dx'}{4D_0\sqrt{\pi Q}}
e^{-(\xi-\xi')^2/Q},\hskip0.7in\cr
j(x,t)&=&\frac{2}{\beta m}\int\frac{p_0(x')dx'}{8D_0\sqrt{\pi Q^3}}
\left(1-e^{-2\tau}\right)\cr &&\hskip0.9in\times
\left(\xi-\xi'\right) e^{-(\xi-\xi')^2/Q}.
\end{eqnarray}
It is easy to check that the mass conservation (\ref{2.3}),
$\partial_{\tau}p+\sqrt{\beta m/2}\partial_{\xi}j=0$ in the scaled
coordinates, is satisfied. The quantity $Q$ plays the role of a
"stretched" time \cite{stretch}. For short times, $t\ll 2m/\gamma$,
$Q\simeq \tau^2$ and $Z\simeq 4\tau^3/3$. The formulas
(\ref{3.5}) and (\ref{3.7}) describe correctly behavior of the
Newtonian particles in this limit. The mapping procedure, as
outlined in the previous Section, requires us to study asymptotic
behavior in the opposite limit, $t\gg 2m/\gamma$.

For large times, $\tau\rightarrow\infty$, the stretched time $Q$
becomes $\tau$ and the formulas (\ref{3.7}) represent the general
solution of the diffusion equation, as expected according to the
previous Section. Now it is necessary to verify that the mass $m$
can serve as a small parameter controlling the series of corrections
to the diffusion equation and its solution.

It may look problematic at first glance, because $Q$ contains
$\exp(-2\tau)=\exp(-\gamma t/m)$, representing essential
singularity of the variable $m\rightarrow 0$. Then the formulas
(\ref{3.7}) [and similarly Eq. (\ref{3.5})] are not expandable in
$m$. Nevertheless, this property is still consistent with the
general scheme of the mapping, as analyzed in Ref. \cite{exact}.

The dimensional reduction, as demonstrated on anisotropic diffusion
in a narrow channel \cite{exact}, also reduces the Hilbert space of
the full-space problem. Let us denote $\hat M(\epsilon)$ the spatial
operator of the evolution equation, i.e. $\hat M(\epsilon)=\partial_x^2
+(1/\epsilon)\partial_y^2$ for anisotropic 2D diffusion; the
eigenvalues $\lambda_i$ and the eigenfunctions $\psi_i(x,y)$ are
given by the equation
\begin{equation}\label{3.8}
-\hat M(\epsilon)\psi_i(\epsilon;x,y)=\lambda_i(\epsilon)
\psi_i(\epsilon;x,y),
\end{equation}
supplemented by proper boundary conditions at the walls of the channel.
The parameter of anisotropy $\epsilon<1$ splits the spectrum into two
parts, the low-lying states, whose eigenvalues $\lambda_l(\epsilon)$
remain finite for $\epsilon\rightarrow 0$ and the transients with
$\lambda_r(\epsilon)$ diverging $\sim 1/\epsilon$. Then
the exact 2D density $\rho$ evolves as
\begin{equation}\label{3.9}
\rho(\epsilon;x,y,t)=\sum_{i}c_i\psi_i(\epsilon;x,y)
e^{-\lambda_i(\epsilon) t},
\end{equation}
the constants $c_i$ are given by the initial condition and the summation
runs over the whole spectrum. The transients contribute to the sum by the
terms proportional to $\exp(-\bar\lambda_r t/\epsilon)$, where
$\bar\lambda_r=\epsilon\lambda_r(\epsilon)$ are finite in the limit
$\epsilon\rightarrow 0$. The result is a formula containing the essential
singularity in the parameter $\epsilon$ near zero, similar to Eq.
(\ref{3.5}) with singular $\exp(-\gamma t/m)$ for $m\rightarrow 0$.

The mapping procedure reduces the full Hilbert space of all $\psi_i$
onto the space defined only by the low-lying states $\psi_l$. If the
1D density $p(x,t)$ is integrated from Eq. (\ref{3.9}) and mapped
backward onto the full Hilbert space (by some operator $\hat\omega$),
the transients will be canceled; the sum (\ref{3.9}) after the mapping
there and back runs only over the low-lying states $\psi_l$. The
terms retained involve no essential singularity in $\epsilon$; the
formula for $\rho$ considered in the mapping procedure represents the
regular part of the exact 2D density $\rho$ with respect to the
parameter $\epsilon$ near zero. This reduction is natural for the
zero-th order (Fick-Jacobs) approximation, as the transients decay
infinitely fast due to their infinite eigenvalues $\lambda_r(\epsilon
\rightarrow 0)$. Nevertheless, the mapping based on fixing
the series of corrections expanded in $\epsilon$ can work only with
the regular part of the 2D density.

Correspondingly, the formulas (\ref{3.5}) and (\ref{3.7}) are exact,
including the contributions of the transients, which are represented
by the singular terms $\sim\exp(-\gamma t/m)$. The mapping requires
us to analyze only the regular parts,
\begin{eqnarray}\label{3.10}
p_{reg}(x,t)&=&\int\frac{p(x')dx'}{2\sqrt{\pi D_0(t-D_0\beta m)}}\cr
&&\times\exp\left[-(x-x')^2)/4D_0(t-D_0\beta m)\right],\hskip0.2in\cr
j_{reg}(x,t)&=&\int\frac{(x-x')p(x')dx'}{4\sqrt{\pi D_0
(t-D_0\beta m)^3}}\cr&&\times\exp\left[-(x-x')^2/
4D_0(t-D_0\beta m)\right]
\end{eqnarray}
and
\begin{eqnarray}\label{3.11}
\rho_{reg}(x,v,t)&=&\sqrt{\frac{\beta m}{2\pi}}\int\frac{p(x')dx'}
{2\sqrt{\pi D_0(t-3D_0\beta m/2)}}\hskip0.4in\cr
&&\hskip-0.5in\times\exp\left[-\frac{(x-x'-D_0\beta mv)^2}{4D_0
(t-3D_0\beta m/2)}-\frac{1}{2}\beta mv^2\right],
\end{eqnarray}
written in the unscaled coordinates, obtained after taking only the
regular parts of $Q$ and $Z$ (\ref{3.6}), $Q_{reg}=\tau-1/2$
and $Z_{reg}=\tau-3/4$, in Eqs. (\ref{3.7}) and (\ref{3.5}). Of course,
the formulas are applied for $t\gg D_0\beta m$, when the transients
vanish. We omit writing the subscript "reg" in the following text.

In comparison with the overdamped limit $m\rightarrow 0$, evolution of
the spatial (1D) density $p$ and the flux $j$ is only corrected by a
time shift, $t\rightarrow t-D_0\beta m$ in the formulas (\ref{3.10}).
The Gaussian distribution is retarded by the time $t_0=D_0\beta m=
m/\gamma$, corresponding to the mean time necessary for losing
information about the original velocity. The value of the shift $t_0$
is constant in $x$ and $t$, so the density $p$ (\ref{3.10}) still
obeys the diffusion equation,
\begin{equation}\label{3.12}
\partial_tp(x,t)=D_0\partial_x^2p(x,t),
\end{equation}
and $j(x,t)=-D_0\partial_xp(x,t)$. There is no correction of the
zero-th order mapped equation due to the finite mass of the particle.
Let us stress that the case of $U(x)=0$ is extremely simple, similar
to the mapping of the 2D diffusion in a flat narrow channel, also giving
no corrections to the Fick-Jacobs approximation.

Nevertheless, the relation of the backward mapping, generating the
phase-space density $\rho$ from the mapped quantities $p$ and $j$, is
not quite trivial. In the formula for $\rho$, Eq. (\ref{3.11}), the
time is shifted by $3t_0/2$ and the displacement $x-x'$ by $vt_0$ with
respect to the distribution of a massless particle. The shortest
way to construct the relation of backward mapping is by applying the
shift operators in $t$ and $x$ on $p(x,t)$, compensating the
different shifts of time and displacement in Eqs. (\ref{3.10})
and (\ref{3.11}),
\begin{equation}\label{3.13}
\rho(x,v,t)=\sqrt{\frac{\beta m}{2\pi}}e^{-\beta mv^2/2}
e^{-(t_0/2)\partial_t-vt_0\partial_x}p(x,t).
\end{equation}
Due to Eq. (\ref{3.12}), the operator $\partial_t$ in the exponent
can be replaced by $D_0\partial_x^2$. After expanding the shift
operator $\exp(-vt_0\partial_x)$ and using $j=-D_0\partial_xp$,
we arrive at the relation of the form (\ref{2.9}),
\begin{eqnarray}\label{3.14}
\rho(x,v,t)&=&\sqrt{\frac{\beta m}{2\pi}}e^{-\beta mv^2/2-(D_0t_0/2)
\partial_x^2}\sum_{k=0}^{\infty}(\beta mv^2)^k\hskip0.3in\cr
&&\hskip-0.5in\times(D_0t_0)^k\partial_x^{2k}\left[\frac{1}{(2k)!}
p(x,t)+\frac{\beta mv}{(2k+1)!}j(x,t)\right];
\end{eqnarray}
the explicit formulas for $\hat\omega$ and $\hat\eta$ to be
substituted in Eq. (\ref{2.9}) in the case $U(x)=0$ are
\begin{eqnarray}\label{3.15}
\hat\omega(x,v)&=&e^{-(D_0t_0/2)\partial_x^2}\sum_{k=0}^{\infty}
\frac{(\beta mv^2)^k}{(2k)!}(D_0t_0)^k\partial_x^{2k},\hskip0.3in\cr
\hat\eta(x,v)&=&e^{-(D_0t_0/2)\partial_x^2}\sum_{k=0}^{\infty}
\frac{(\beta mv^2)^k}{(2k+1)!}(D_0t_0)^k\partial_x^{2k}.
\end{eqnarray}
Both operators are expandable in $m$, ($t_0=m/\gamma$); their zero-th
order coefficients equal unity, consistent with the heuristic formula
(\ref{2.6}). Also applying the relation (\ref{3.14}) in the definitions
(\ref{2.1}) and (\ref{2.4}) gives identity.

The final step is that of verifying the evolution equation for $j$, Eq.
(\ref{2.5}). Two integrals are to be completed with $\rho$ expressed
by the backward mapping, Eq. (\ref{3.14}),
\begin{eqnarray}\label{3.16}
\int_{-\infty}^{\infty}v^2\rho(x,y,t)dv=\left(1+D_0t_0\partial_x^2
\right)p(x,t)/\beta m,\cr
\int_{-\infty}^{\infty}e^{-\beta mv^2/2}\partial_ve^{\beta mv^2/2}
\rho(x,v,t)dv=\beta m j(x,t),\ \ 
\end{eqnarray}
which are now valid exactly. If substituted in Eq. (\ref{2.5}), and
the equation (\ref{3.12}) is applied, we get
\begin{equation}\label{3.17}
(1+t_0\partial_t)[j(x,t)+D_0\partial_xp(x,t)]=0
\end{equation}
after simple algebra. This equation validates the relation
$j=-D_0\partial_xp$ for nonzero $m$ as well and thus, if combined with
the mass conservation (\ref{2.3}), also the diffusion equation
(\ref{3.12}) without any corrections. If compared with the calculation
of Eq. (\ref{2.8}), the 1-st order term  $\sim\partial_tj$, neglected
in the previous Section, is compensated here by other 1-st order term
coming from the exact relation (\ref{3.14}), appearing in the integrals
(\ref{3.16}).

The FP equation (\ref{1.2}) with zero potential is too simple to give
nonzero corrections to the diffusion equation (\ref{3.12}).
Nevertheless, it helped us to understand the structure of the mapping.
It shows that the scheme suggested in the previous section is viable.
The relation of the backward mapping has the form of Eq. (\ref{2.9}),
at least for $U(x)=0$, and the operators $\hat\omega$ and $\hat\eta$
can be expanded in $m$, or $t_0=m/\gamma$,
\begin{equation}\label{3.18}
\hat\omega=\sum_{k=0}^{\infty}t_0^k\hat\omega_k(x,u),\hskip0.2in
\hat\eta=\sum_{k=0}^{\infty}t_0^k\hat\eta_k(x,u).
\end{equation}
Integration in Eqs. (\ref{3.16}) indicates that the coefficients
$\hat\omega_k$ and $\hat\eta_k$ should be sought dependent up on the
scaled velocity $u$ rather than $v$ (compare to Ref. \cite{Hot});
otherwise each term in Eq. (\ref{3.18}) would contribute in several
succeeding orders in the integrals (\ref{3.16}).
The mixing of orders would hinder us in constructing the recurrence
scheme generating corrections to Eq. (\ref{2.5}). In the notation
of Eq. (\ref{3.18}), $\hat\omega_0=\hat\eta_0=1$ and
\begin{eqnarray}\label{3.19}
\hat\omega_1(x,u)&=&(u^2-1/2)D_0\partial_x^2,\hskip0.3in\cr
\hat\eta_1(x,u)&=&(u^2/3-1/2)D_0\partial_x^2,\cr
\hat\omega_2(x,u)&=&(u^4/6-u^2/2+1/8)D_0^2\partial_x^4,\cr
.&.&.
\end{eqnarray}
valid for $U(x)=0$ according to Eqs. (\ref{3.15}), will be used for
testing the results of the recurrence procedure in the next Section.

\renewcommand{\theequation}{4.\arabic{equation}}
\setcounter{equation}{0}

\section{IV. Mapping procedure}

We now finish the construction of the mapping procedure, outlined in
the Section II, for an arbitrary analytic potential $U(x)$. Supposing
the backward mapping of the form (\ref{2.9}) with the operators
$\hat\omega$ and $\hat\eta$ expanded in $t_0$ ($m$) according to
Eqs. (\ref{3.18}), we find recurrence relations fixing the coefficients
$\hat\omega_k$ and $\hat\eta_k$. Completing the integrals in Eq.
(\ref{2.5}), we obtain a series of corrections to the zero-th order
relation $j=-D_0\partial_xp$. Combined with mass conservation,
Eq. (\ref{2.3}), it gives the Smoluchowski equation corrected due to
nonzero mass of the particle.

The essential relation determining the operators $\hat\omega$ and
$\hat\eta$ is the FP equation (\ref{1.2}), which has to be satisfied
for any solution of the reduced problem, the density $p(x,t)$ and
the flux $j(x,t)$, after their backward mapping (\ref{2.9}) onto the
full-dimensional Hilbert space. If the expansion in $t_0=m/\gamma$
of both operators, Eq. (\ref{3.18}), is supposed, we have
\begin{eqnarray}\label{4.1}
&&\hskip-0.3in\left[\partial_t+\sqrt{\frac{2}{\beta m}}u\partial_x
-\frac{\beta U'(x)}{\sqrt{2\beta m}}\partial_u-\frac{1}{2t_0}
\partial_u e^{-u^2}\partial_u e^{u^2}\right]\sum_{k=0}^{\infty}
t_0^k\cr &&\hskip-0.3in\times e^{-u^2}\left[\hat\omega_k
(x,u)p(x,t)+\sqrt{2\beta m}u\hat\eta_k(x,u)j(x,t)\right]=0
\end{eqnarray}
after introducing the scaled velocity $u=\sqrt{\beta m/2}v$ in Eq.
(\ref{1.2}). The factor $\beta m$ is replaced by $t_0/D_0$ in the
following calculations. Thus half-integer powers of $t_0$ appear
in Eq. (\ref{4.1}) \cite{Hot}. As this equation has to be satisfied
for any $t_0$, we can split it for clarity into two relations: the first
one, including only the integer powers of $t_0$,
\begin{eqnarray}\label{4.2}
\sum_{k=0}^{\infty}t_0^k\bigg[\partial_t\hat\omega_kp
+2u^2\partial_x\hat\eta_kj-\beta U'(x)e^{u^2}\partial_u
ue^{-u^2}\hat\eta_kj\cr -\frac{1}{2t_0}e^{u^2}\partial_u
e^{-u^2}\partial_u\hat\omega_kp\bigg]=0,\ \ \ 
\end{eqnarray}
and the second one, collecting the half-integer powers,
\begin{eqnarray}\label{4.3}
\sum_{k=0}^{\infty}t_0^{k-1/2}\bigg[2t_0u\partial_t\hat\eta_kj
+D_0\Big(2u\partial_x-\beta U'(x)e^{u^2}\partial_ue^{-u^2}\Big)\cr
\times\hat\omega_kp-e^{u^2}\partial_u e^{-u^2}\partial_u u
\hat\eta_k j\bigg]=0.\hskip0.3in
\end{eqnarray}

Notice that Eqs. (\ref{4.2}) and (\ref{4.3}) do not violate parity
of $\hat\omega_k$ and $\hat\eta_k$ in $u$. If used for construction
of the recurrence relations between the coefficients, all they have
to have the same parity as $\hat\omega_0=\hat\eta_0=1$; hence
$\hat\omega_k(x,u)=\hat\omega_k(x,-u)$ and $\hat\eta_k(x,u)=
\hat\eta_k(x,-u)$. This symmetry enables us to find the normalization
(or identity) conditions for $\hat\omega_k$ and $\hat\eta_k$. The
backward mapped $\rho$, Eq. (\ref{2.9}), with the operators
$\hat\omega$, $\hat\eta$ expanded in $t_0$, Eq. (\ref{3.18}),
substituted in the definitions (\ref{2.1}) and (\ref{2.4}) has to give
identities for any $t_0$, $p(x,t)$ and $j(x,t)$. Thus we obtain
\begin{equation}\label{4.4}
\frac{1}{\sqrt{\pi}}\int_{-\infty}^{\infty}du e^{-u^2}
\hat\omega_k(x,u)=\delta_{0,k},
\end{equation}
\begin{equation}\label{4.5}
\frac{1}{\sqrt{\pi}}\int_{-\infty}^{\infty}2u^2du e^{-u^2}
\hat\eta_k(x,u)=\delta_{0,k}.
\end{equation}

The operators $\hat\omega_k$ and $\hat\eta_k$ are supposed not to depend
on time, so the time derivative commutes with them and acts directly on
$p(x,t)$ or $j(x,t)$ in Eqs. (\ref{4.2}), (\ref{4.3}). To derive the
operators unambiguously, using only spatial derivatives, we express
$\partial_tp=-\partial_xj$ from the mass conservation, Eq. (\ref{2.3}).
However, the time derivative of $j$ cannot be expressed in a similar way
from Eq. (\ref{2.5}), because $\partial_tj$ is not the leading term there.
If the backward mapping, Eqs. (\ref{2.9}) and (\ref{3.18}), is applied
to the integrals of Eq. (\ref{2.5}), we get
\begin{equation}\nonumber
\int_{-\infty}^{\infty}v^2\rho(x,v,t) dv=\frac{1}{\beta m}\left(1+
\sum_{k=1}^{\infty}t_0^k\hat I_k(x)\right)p(x,t),
\end{equation}
where the operators $\hat I_k$ are given by
\begin{equation}\label{4.6}
\hat I_k(x)=\frac{2}{\sqrt{\pi}}\int_{-\infty}^{\infty}
u^2du e^{-u^2}\hat\omega_k(x,u);
\end{equation}
the right-hand side integral of Eq. (\ref{2.5}) results in
\begin{eqnarray}\nonumber
\frac{\gamma}{\beta m^2}\int_{-\infty}^{\infty}e^{-\beta mv^2/2}
\partial_v e^{\beta mv^2/2}\rho(x,v,t)dv&=&\cr
\frac{\gamma}{\sqrt{\pi}m}\sum_{k=0}^{\infty}t_0^k
\int_{-\infty}^{\infty}due^{-u^2}\partial_uu\hat\eta_k(x,u)
j(x,t)&=&\frac{j(x,t)}{t_0}
\end{eqnarray}
after integrating by parts and using the normalization relation (\ref{4.5}).
Then, instead of the evolution equation for $j$, Eq. (\ref{2.5})
has to be understood as an expression relating $j$ and $p$,
\begin{eqnarray}\label{4.7}
(1+t_0\partial_t)j(x,t)&=&-D_0\bigg[e^{-\beta U(x)}
\partial_x e^{\beta U(x)}\hskip0.7in\cr &&\hskip0.3in
+\partial_x\sum_{k=1}^{\infty}t_0^k\hat I_k(x)\bigg]p(x,t),
\end{eqnarray}
and the flux $j$, as well as its time derivative, is expressed using
$p$ according to this equation. The term $t_0\partial_t$ acts now on $p$
as the 1-st order correction in $(1+t_0\partial_t)^{-1}p$ after completing
inversion and commutation with the spatial operators in the square brackets
of Eq. (\ref{4.7}). Then $\partial_tp$ is systematically replaced by
$-\partial_x j$, contributing to the next corrections in the higher orders
of $t_0$. The result is a formula for $j$ expressed by some purely spatial
operator acting on $p$,
\begin{equation}\label{4.8}
j(x,t)=-D_0 e^{-\beta U(x)}\left[1+\sum_{k=1}^{\infty}t_0^k
\hat Z_k(x)\right]\partial_x e^{\beta U(x)}p(x,t),
\end{equation}
where the operators $\hat Z_k$ are related to $\hat I_k$, Eq. (\ref{4.6}).
Using this form, we are able to write the final mapped equation explicitly
after combination with Eq. (\ref{2.3}),
\begin{eqnarray}\label{4.9}
\partial_tp&=&D_0\partial_x e^{-\beta U(x)}\left[1+\sum_{k=1}^{\infty}t_0^k
\hat Z_k\right]\partial_x e^{\beta U(x)}p\cr
&=&\sum_{k=0}^{\infty}t_0^k\hat Q_kp=\hat Qp,
\end{eqnarray}
which is the Smoluchowski equation (\ref{1.3}) in the zero-th order,
extended by a series of mass dependent corrections in $t_0$. The operators
$\hat Q_k$ are introduced to simplify notation in the following calculations.

Now the operators $\hat Z_k$ can be expressed explicitly using $\hat I_k$.
Expanding $(1+t_0\partial_t)^{-1}p$ in $t_0$ and applying Eq. (\ref{4.9}),
we have
\begin{equation}\nonumber
(1+t_0\partial_t)^{-1}p=(1-t_0\hat Q+...)p=\Big(1+
\sum_{k=0}^{\infty}t_0^{k+1}\hat Q_k\Big)^{-1}\hskip-0.1in p,
\end{equation}
which is to be used in the operator equation,
\begin{eqnarray}\label{4.10}
&&\hskip-0.2in e^{-\beta U(x)}\left[1+\sum_{k=1}^{\infty}t_0^k
\hat Z_k\right]\partial_xe^{\beta U(x)}=\bigg[e^{-\beta U(x)}
\partial_x e^{\beta U(x)}\cr
&&\hskip0.6in+\partial_x\sum_{k=1}^{\infty}t_0^k\hat I_k\bigg]
\Big(1+\sum_{k=0}^{\infty}t_0^{k+1}\hat Q_k\Big)^{-1},
\end{eqnarray}
obtained after comparison of Eqs. (\ref{4.7}) and (\ref{4.8}). Expanding
in $t_0$ and comparing the coefficients of the same powers of $t_0$,
we get a sequence of relations fixing $\hat Z_k$,
\begin{eqnarray}\label{4.11}
e^{-\beta U}\hat Z_1\partial_x e^{\beta U}&=&
\partial_x \hat I_1-e^{-\beta U}\partial_x e^{\beta U}\hat Q_0,\cr
e^{-\beta U}\hat Z_2\partial_x e^{\beta U}&=&
\partial_x \hat I_2-\partial_x \hat I_1\hat Q_0\cr
&&+e^{-\beta U}\partial_x e^{\beta U}(\hat Q_0^2-\hat Q_1),\cr
.&.&.
\end{eqnarray}

Finally, we derive the recurrence scheme, generating the operators
$\hat\omega_k$, determining $\hat I_k$, Eq. (\ref{4.6}), and
thus also $\hat Z_k$ or $\hat Q_k$, Eqs. (\ref{4.9}), (\ref{4.11}).
Notice that the $\hat\eta_k$ do not directly enter the mapped equation
(\ref{4.9}), but they are necessary in the recurrence formulas for
$\hat\omega_k$.

The recurrence scheme for $\hat\omega_k$ is defined by Eq. (\ref{4.2}).
This equation has to be satisfied for any $p$ and $j$, solving the
mapped problem, but these quantities, although considered before as
formally independent, are related by Eq. (\ref{4.8}). To obtain an
equation for operators, $j$ has to be expressed by $p$. Applying the
relation (\ref{4.8}) and Eq. (\ref{4.9}) for $\partial_tp$, we get
\begin{eqnarray}\label{4.12}
&&\hskip-0.2in\sum_{n=0}^{\infty}t_0^n\bigg[D_0\Big(\hat\omega_n\partial_x-
2u^2\partial_x\hat\eta_n+\beta U'e^{u^2}\partial_uu
e^{-u^2}\hat\eta_n\Big)e^{-\beta U(x)}\cr
&&\hskip-0.1in\times\sum_{l=0}^{\infty}
t_0^l\hat Z_l\partial_xe^{\beta U(x)}-\frac{1}{2t_0}e^{u^2}
\partial_ue^{-u^2}\partial_u\hat\omega_n\bigg]p=0,
\end{eqnarray}
valid for any function $p(x,t)$; we have $\hat Z_0=1$. To lowest
order, $t_0^{-1}$, only the term $e^{u^2}\partial_ue^{-u^2}\partial_u
\hat\omega_0=0$. It is satisfied by $\hat\omega_0=1$, the only
solution nondiverging at $u\rightarrow\pm\infty$ and also satisfying
the normalization, Eq. (\ref{4.4}). In the higher orders, we derive
the recurrence relation,
\begin{eqnarray}\label{4.13}
e^{u^2}\partial_ue^{-u^2}\partial_u\hat\omega_{n+1}&=&2D_0
\sum_{k=0}^n\Big(\hat\omega_k\partial_x-2u^2\partial_x\hat\eta_k
+\beta U'(x)\cr
&&\hskip-0.8in \times e^{u^2}\partial_uu e^{-u^2}\hat\eta_k\Big)
e^{-\beta U(x)}\hat Z_{n-k}\partial_xe^{\beta U(x)}.
\end{eqnarray}

Calculation of the $\hat\omega_{n+1}$ requires us to know the $\hat\eta_k$
up to $k=n$. They are generated from Eq. (\ref{4.3}). Again, $j$, as well
as $\partial_t j$, have to be expressed by $p$ using the relations
(\ref{4.8}) and (\ref{4.9}). Then Eq. (\ref{4.3}) becomes
\begin{eqnarray}\label{4.14}
&&\sum_{n=0}^{\infty}t_0^{n-1/2}\bigg[
e^{u^2}\partial_u e^{-u^2}\partial_u u\hat\eta_n e^{-\beta U}
\sum_{k=0}^{\infty}t_0^k\hat Z_k\partial_xe^{\beta U}\cr
&&-2D_0t_0u\hat\eta_ne^{-\beta U}
\sum_{k=0}^{\infty}t_0^k\hat Z_k\partial_xe^{\beta U}\partial_x
e^{-\beta U}\sum_{l=0}^{\infty}t_0^l\hat Z_l\partial_xe^{\beta U}\cr
&&+\big(2u\partial_x-\beta U'e^{u^2}\partial_u e^{-u^2}\big)
\hat\omega_n\bigg]p=0
\end{eqnarray}
valid for any function $p(x,t)$. In lowest order, $t_0^{-1/2}$, 
\begin{eqnarray}\label{4.15}
e^{u^2}\partial_ue^{-u^2}\partial_u u\hat\eta_0e^{-\beta U}\partial_x
e^{\beta U}&=&-2u\partial_x+\beta U'e^{u^2}\partial_ue^{-u^2}\cr
=-2u(\partial_x+\beta U')&=&-2ue^{-\beta U}\partial_x e^{\beta U},
\end{eqnarray}
where we have used $\hat\omega_0=1$ and $\hat Z_0=1$. After the first
integration, we have
\begin{equation}\label{4.16}
\partial_uu\hat\eta_0=-e^{u^2}\int 2udue^{-u^2}=1+\hat C_1e^{u^2};
\end{equation}
the integration constant $\hat C_1=0$ provides convergence as
$u\rightarrow\pm\infty$. The next integration gives $\hat\eta_0
=1+(1/u)\hat C_0$; $\hat C_0=0$. This calculation validates our
heuristic formula (\ref{2.6}) in the zero-th order approximation.

In the higher orders, Eq. (\ref{4.14}) generates the relations
\begin{eqnarray}\label{4.17}
&&\hskip-0.2in e^{u^2}\partial_ue^{-u^2}\partial_u u\hat\eta_n
e^{-\beta U}\partial_xe^{\beta U}=(\beta U'e^{u^2}\partial_u
e^{-u^2}-2u\partial_x)\hat\omega_n\cr
&&+2D_0\sum_{k,l=0}^{k+l<n} u\hat\eta_{n-k-l-1}
e^{-\beta U}\hat Z_k\partial_x e^{\beta U}\partial_x e^{-\beta U}
\hat Z_l\partial_x e^{\beta U}\cr
&&-\sum_{k=0}^{n-1}e^{u^2}\partial_ue^{-u^2}\partial_u
u\hat\eta_k e^{-\beta U}\hat Z_{n-k}\partial_xe^{\beta U},
\end{eqnarray}
forming the recurrence scheme for $\hat\eta_n$. Completing the
operations in Eqs. (\ref{4.17}) and (\ref{4.13}) one has to keep
in mind that the equation acts on an arbitrary function $p(x,t)$,
not depending on $u$. On the other hand, the operators $\hat\omega_k$
and $\hat\eta_k$ for $k>0$ depend on $u$.

The recurrence procedure starts from $\hat\omega_0=1$ and $\hat Z_0=1$.
Calculation of the next order correction requires first expressing the
$\hat\eta_n$ according to Eq. (\ref{4.17}), or (\ref{4.15}) for $n=0$,
as shown above. Then $\hat\omega_{n+1}$ is derived from Eq. (\ref{4.13}),
$\hat I_{n+1}$ integrated according to Eq. (\ref{4.6}) and finally
$\hat Z_{n+1}$ expressed from Eq. (\ref{4.11}). To demonstrate the
procedure, we derive the first order correction, $\hat Z_1$.

We use already calculated $\hat\eta_0=1$. For $n=0$, Eq. (\ref{4.13})
becomes
\begin{equation}\label{4.18}
e^{u^2}\partial_u e^{-u^2}\partial_u\hat\omega_1=
2D_0(1-2u^2)e^{-\beta U(x)}\partial_x^2 e^{\beta U(x)}.
\end{equation}
After the first integration,
\begin{equation}\label{4.19}
\partial_u\hat\omega_1=2D_0e^{u^2}\left(ue^{-u^2}+C_1\right)
e^{-\beta U(x)}\partial_x^2 e^{\beta U(x)},
\end{equation}
the integration constant $C_1=0$, to provide convergence for
$u\rightarrow\pm\infty$. The integration constant $C_0$ after the
next integration is fixed to satisfy the normalization, Eq.
(\ref{4.4}),
\begin{equation}\label{4.20}
\int_{-\infty}^{\infty}due^{-u^2}D_0(u^2+C_0)
e^{-\beta U(x)}\partial_x^2 e^{\beta U(x)}=0,
\end{equation}
hence $C_0=-1/2$ and
\begin{equation}\label{4.21}
\hat\omega_1=D_0(u^2-1/2)e^{-\beta U(x)}\partial_x^2 e^{\beta U(x)}.
\end{equation}
For $U(x)=0$, we recover the corresponding formula in Eq. (\ref{3.19}).
Integration over $u$ in Eq. (\ref{4.6}) results in
$\hat I_1=D_0\exp[-\beta U(x)]\partial_x^2\exp[\beta U(x)]$,
giving finally
\begin{equation}\label{4.22}
\hat Z_1=D_0\left(e^{\beta U}\partial_x e^{-\beta U}\partial_x-
\partial_x e^{\beta U}\partial_xe^{-\beta U}\right)=D_0\beta U''(x)
\end{equation}
from Eq. (\ref{4.11}).

In the higher orders, $\hat\eta_n$, $\hat\omega_{n+1}$ are calculated
according to Eqs. (\ref{4.17}) and (\ref{4.13}). The integration constants
after double integration have to provide convergence for $u\rightarrow
\pm\infty$, requiring the operators to be even in $u$, and the
normalization, Eq. (\ref{4.4}). The condition (\ref{4.5}) for
$\hat\eta_n$ is satisfied automatically; it serves as a check on the
computation. The derivation is tedious, and we present only the results
in second order,
\begin{eqnarray}\label{4.23}
\hat\eta_1&=&D_0(u^2/3-1/2)e^{-\beta U(x)}\partial_x^2
e^{\beta U(x)},\cr
\hat\omega_2&=&\frac{D_0^2}{2}e^{-\beta U(x)}\Big[\Big(
\frac{u^4}{3}-u^2+\frac{1}{4}\Big)\partial_x^3\cr
&&+\Big(u^2-\frac{1}{2}\Big)\big(4\beta U''(x)\partial_x
+3\beta U^{(3)}(x)\big)\Big]\partial_x e^{\beta U(x)},\cr
\hat Z_2&=&\frac{D_0^2}{2}\Big[4\big(\beta U''(x)\big)^2-
\beta^2U'(x)U^{(3)}(x)+\beta U^{(4)}(x)\cr
&&\hskip0.3in +3\beta U^{(3)}(x)\partial_x\Big].
\end{eqnarray}

Again, the formulas (\ref{3.19}) for $U(x)=0$ are recovered. There
are no contributions to $\hat Z_n$ in this case, too, as expected
according to the analysis in the previous Section. Also, linear
potentials, $U(x)=-Fx$, have no effect on validity of the
uncorrected Smoluchowski equation (\ref{1.3}). The particle
driven by a constant force $F$ move asymptotically with constant
mean velocity $v_0=F/\gamma$ and the distribution $p(y,t)$ in
coordinate $y$, shifted by the drift, $y=x-v_0t$, is again Gaussian
as in the case of no potential.

The situation becomes different if the driving force $F(x)$ is not
constant. If the mass $m$ or the time of the thermalization
$t_0=m/\gamma$ is small, but nonzero, the particle appearing at a
new position $x$ has to accommodate to the new local mean velocity.
On the other hand, it carries some mean momentum from its previous
position and needs some time to change it.
Meanwhile it slips to some other position than predicted
by purely stochastic dynamics due to its inertia, or non-zero mass.
The effects of such slipping are indicated by the corrections
$\hat Z_n$ of the Smoluchowski equation and they are nonzero for
potentials with nonzero $U''(x)$, or higher derivatives.

As seen from Eq. (\ref{4.23}), the $\hat Z_n$ are not only functions,
but operators, containing $\partial_x$ in the higher orders. So the
mapped equation (\ref{4.9}) has exactly the same structure as the
mapped equations for diffusion \cite{eff}, or biased diffusion
\cite{soft, gravi, forced}. Being inspired by these works, Eq.
(\ref{4.9}) can be simplified by replacing the correction operators
$D_0[1+\sum_{k=1}^{\infty}t_0^k\hat Z_k]$ by a function $D(x)$, a
spatially dependent effective diffusion coefficient,
\begin{equation}\label{4.24}
\partial_tp(x,t)=\partial_xe^{-\beta U(x)}D(x)\partial_x
e^{\beta U(x)}p(x,t),
\end{equation}
which becomes valid in the limit of stationary flow, i.e. when the
spatial density $p$ and the flux $j$ change very slowly, $p(x,t)
\rightarrow p(x)$. Due to mass conservation, Eq. (\ref{2.3}),
the flux $j(x,t)=j$ is constant (but nonzero) in $x$ as well.
If expressed from Eq. (\ref{4.24}),
\begin{equation}\label{4.25}
j=-e^{-\beta U(x)}D(x)\partial_xe^{\beta U(x)}p(x),
\end{equation}
the function $\partial_x(\exp[\beta U(x)]p(x))=-j\exp[\beta U(x)]/D(x)$
is dependent only on the system; $m$, $\gamma$ and the potential
$U(x)$, for any stationary solution $p(x)$. So we can substitute for
it in Eq. (\ref{4.8}),
\begin{equation}\label{4.26}
j=D_0e^{-\beta U(x)}\left[1+\sum_{k=1}^{\infty}t_0^k\hat Z_k(x)\right]
e^{\beta U(x)}\frac{j}{D(x)},
\end{equation}
valid for stationary flow, and calculate $D(x)$ unambiguously
from the expansion of the corrections $\hat Z_k$,
\begin{equation}\label{4.27}
\frac{D_0}{D(x)}=e^{-\beta U(x)}\left[1+\sum_{k=1}^{\infty}t_0^k
\hat Z_k(x)\right]^{-1}e^{\beta U(x)},
\end{equation}
as a series in $t_0$,
\begin{eqnarray}\label{4.28}
D(x)/D_0&=&1+D_0t_0\beta U''+(D_0t_0)^2\Big[2(\beta U'')^2\cr
&&\hskip0.2in+\beta^2U'U^{(3)}+\beta U^{(4)}/2\Big]+...\ .
\end{eqnarray}

The next simplification is that of neglecting all the derivatives but
$U''(x)$, i.e. approximating the real $U(x)$ locally by a quadratic potential.
In this case, the expansion (\ref{4.28}) can be summed up to infinity,
\begin{eqnarray}\label{4.29}
D(x)/D_0&\simeq&\sum_{n=0}^{\infty}
\frac{(2n)!}{n!(n+1)!}\left[D_0t_0\beta U''(x)\right]^n\cr
&&=\frac{1-\sqrt{1-4D_0t_0\beta U''(x)}}{2D_0t_0\beta U''(x)},
\end{eqnarray}
the proof is given in the Appendix B. The formula works for
$D_0t_0\beta U''(x)=mU''(x)/\gamma^2<1/4$, which is the condition
for non-oscillatory movement of a particle in a quadratic well with
friction, the damped harmonic oscillator. If $U''(x)=\kappa$ is
constant, the trajectory of a particle averaged over the stochastic
force is governed by
\begin{equation}\label{4.30}
m\langle\ddot x\rangle+\gamma\langle\dot x\rangle +\kappa
\langle x\rangle=0
\end{equation}
from Eq. (\ref{1.1}). The particular solutions are
$\langle x(t)\rangle=\exp(\alpha t)$ with $\alpha=-(\gamma\pm
\sqrt{\gamma^2-4m\kappa})/2m$. Requiring $\alpha$ to be a real number
gives the same condition.

This simple example demonstrates restriction of the theory presented
to nonoscillatory movement of the particle in potential wells on
its way along a 1D channel. A small mass $m$ is expected, to enable
the friction quickly to damp the momentum of a particle; i.e. to have
the relaxation in the velocity faster than in the real space coordinate
$x$.

\begin{figure}
\includegraphics[scale=0.4]{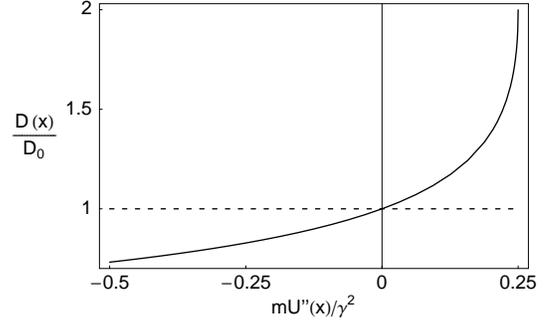}
\caption{Plot of the effective diffusion coefficient $D(x)$ dependent
on the mass of the particle $m$ and friction $\gamma$ according to
Eq. (\ref{4.29}), valid if the derivatives of the potential higher
than $U''(x)$ are neglected.}
\end{figure}

A more detailed insight to the restrictions of the dimensional reduction
of the phase space controlled by the mass $m$ can be obtained by comparison
of the Green's function (GF) of the mapped equation (\ref{4.24}) with $D(x)$
given by Eq. (\ref{4.29}) and GF of the Kramers equation (\ref{1.2}) for
the damped harmonic oscillator, $U(x)=\kappa x^2/2=m\omega_0^2x^2/2$,
which is exactly solvable. The solution $G=G(x,v,t;x',v',t')$ of the equation
\begin{eqnarray}\label{4.31}
\Big(\p_t+v\p_x-\omega_0^2x\p_v-\frac{\gamma}{\beta m^2}\p_v
e^{-\beta mv^2/2}\p_v e^{\beta mv^2/2}\Big) G\cr
=\delta(x-x')\delta(v-v')\delta(t-t')\hskip1cm
\end{eqnarray}
reads \cite{Chand}
\begin{eqnarray}\label{4.32}
G=\frac{(s_1-s_2)e^{\gamma t/m}}{2\pi\sqrt{ab-h^2}}\exp\Big[-\Big(
a(\xi-\xi_0)^2+b(\eta-\eta_0)^2\cr +2h(\xi-\xi_0)(\eta-\eta_0)\Big)
\big/2(ab-h^2)\Big],\hskip1cm
\end{eqnarray}
where
\begin{equation}\label{4.33}
s_{1,2}=-\frac{\gamma}{2m}\pm\sqrt{\frac{\gamma^2}{4m^2}-\omega_0^2},
\end{equation}
\begin{eqnarray}\label{4.34}
\xi=(s_1x-v)e^{-s_2t},\hskip1cm \xi_0=(s_1x'-v'),\hskip1.3cm\cr
\eta=(s_2x-v)e^{-s_1t},\hskip1cm \eta_0=(s_2x'-v'),\hskip1.3cm\cr
a=\frac{\gamma}{\beta m^2s_1}\Big(1-e^{-2s_1t}\Big),\ \ 
b=\frac{\gamma}{\beta m^2s_2}\Big(1-e^{-2s_2t}\Big),\cr 
h=\frac{2}{\beta m}\Big(1-e^{\gamma t/m}\Big).\hskip2cm
\end{eqnarray}

Similar to the case of no potential in the Section III, let us suppose
that a thermalized particle (equilibrated in velocity) was inserted at
a position $x_0$ at time $t'=0$,
\begin{equation}\label{4.35}
\rho_0(x',v')=\sqrt{\frac{\beta m}{2\pi}}\delta(x'-x_0)
e^{-\beta mv'^2/2}.
\end{equation}
After integration over $v'$ and $v$, we get the corresponding spatial
density
\begin{eqnarray}\label{4.36}
p(x,t)=\int_{-\infty}^{\infty}G(x,v,t;x'v',0)\rho_0(x',v')dx'dv'dv=\cr
\sqrt{\frac{\beta m}{2\pi Z}}\omega_0(s_1-s_2)\exp\Big[
-\frac{\beta m\omega_0^2}{2 Z}\hskip1.5cm \cr
\times \Big(s_1(x-x_0e^{s_2t}) -s_2(x-x_0e^{s_1t})\Big)^2\Big];\hskip0.5cm
\end{eqnarray}
\begin{equation}\nonumber
Z=\Big[s_1(1+e^{s_2t})-s_2(1+e^{s_1t})\Big]
\Big[s_1(1-e^{s_2t})-s_2(1-e^{s_1t})\Big].
\end{equation}

On the other hand, the coefficient $D(x)$, Eq. (\ref{4.29}), becomes
constant for the quadratic potential,
\begin{equation}\label{4.37}
D(x)=\frac{\gamma/2m-\sqrt{\gamma^2/4m^2-\omega_0^2}}{\beta m\omega_0^2}=
\frac{-s_1}{\beta m\omega_0^2},
\end{equation}
and GF of the corresponding mapped equation (\ref{4.24}),
\begin{eqnarray}\label{4.38}
\Big(\p_t-\p_xD(x)e^{-\beta m\omega_0^2x^2/2}\p_xe^{\beta m\omega_0^2x^2/2}
\Big)g(x,t;x_0,t_0)\cr =\delta(x-x_0)\delta(t-t_0),\hskip2cm
\end{eqnarray}
can be easily found by a calculation similar to the derivation of
Eq. (\ref{3.1}), Appendix A. The result,
\begin{eqnarray}\label{4.39}
g(x,t;x_0,0)=\sqrt{\frac{\beta m}{2\pi (1-e^{2s_1t})}}\omega_0
\exp\Big[-\frac{1}{2}\beta m\omega_0^2\cr \times\big(x-x_0e^{s_1t}\big)^2
\big/\big(1-e^{2s_1t}\big)\Big],\hskip0.5cm
\end{eqnarray}
describes evolution of the real space density of a thermalized particle
inserted at $x_0$, too, and can be directly compared with the formula
(\ref{4.36}).

First, let us notice that in comparison with Eq. (\ref{4.36}), the
exponential $e^{s_2t}$ disappeared from the formula (\ref{4.39}).
The root $s_2\simeq -\gamma/m$ for $m\rightarrow 0$ makes
$e^{s_2t}\simeq e^{-\gamma t/m}$ the term essentially singular in $m$ and
so invisible for the recurrence procedure, which works with the operators
$\hat\omega$ and $\hat\eta$ expanded in $m$. Using our argumentation
from the Section III, $e^{s_2t}$ represents the "transients" neglected
by the mapping. On the other hand, $e^{s_1t}\simeq e^{-m\omega_0^2t/
\gamma}$ is regular in $m$ small, representing the contribution of the
low-lying states, retained by the method.

Next, let us stress that the equation (\ref{4.24}) with $D(x)$ expressed
by the expansion (\ref{4.28}) was derived in the limit of the stationary
flow; for the net flux almost constant, which is not the case of the
process described by the Eqs. (\ref{4.36}) and (\ref{4.39}).
Nevertheless, using the approximations
\begin{eqnarray}\nonumber
s_1-s_2\big(1\pm e^{s_1t}\big)&\simeq& (s_1-s_2)\big(1\pm e^{s_1t}\big),\cr
s_1x-s_2\big(x-x_0e^{s_1t}\big)&\simeq& (s_1-s_2)\big(x-x_0e^{s_1t}\big),
\end{eqnarray}
applicable for $e^{s_1t}\ll 1$, the regularized formula (\ref{4.36}) (with
$e^{s_2t}$ neglected) becomes finally Eq. (\ref{4.39}); i.e. it represents
correctly the asymptotic behavior of the spatial density $p(x,t)$ for large
time $t$.

If $m$ approaches $\gamma/2\omega_0$, the transients contributing by
$e^{s_2t}$, neglected by the mapping, become important. In the
oscillating regime, $s_{1,2}$ are complex numbers and both are necessary
for expressing the real density $p(x,t)$ in Eq. (\ref{4.36}). The mapping
which splits the Hilbert space to the retained low-lying states and the
neglected transients, controlled by $m$ small, loses its justification
and the method stops working. Mapping in this region requires a different
method to be applied. It will be an object of our study in the future.

\renewcommand{\theequation}{5.\arabic{equation}}
\setcounter{equation}{0}

\section{V. Conclusion}

Although modeling of transport in confined systems is often
based on study of the Langevin equation, Eq. (\ref{1.1}) in the simplest
1D case, the solutions necessary in practical applications are often
accessible only in two limits:
either the friction $\gamma\rightarrow 0$, when the particles obey
Newtonian dynamics, or the mass $m\rightarrow 0$, which corresponds to
stochastic dynamics. Any solution in the region between these limits
requires working in phase space, which makes the problem much more
complicated.

The present paper shows how to describe the region of finite
$m/\gamma$ while still working in real space, as in the case of
stochastic dynamics. The equation governing evolution of the spatial
density $p(x,t)$ is the Smoluchowski equation, corresponding
to the limit of a massless particle, extended by a series of 
corrections in powers of $t_0=m/\gamma$; $t_0$ can be interpreted as
the typical time of thermalizing of the particle's initial velocity
by the stochastic force.

In general, the extended Smoluchowski equation has the form of Eq.
(\ref{4.9}), $D_0=1/\gamma\beta$ denotes the diffusion constant and
the operators $\hat Z_k$ are systematically derived within the
recurrence procedure presented in the Sect. IV. In the limit of
stationary flow, i.e. when the flux is almost constant but nonzero in
time and space, this equation can be simplified to Eq. (\ref{4.24}),
where the effective diffusion coefficient $D(x)$ is calculated
unambiguously from the operators $\hat Z_k$, Eq. (\ref{4.28}). In the
simplest approximation, when all the derivatives of the potential higher
than $U''(x)$ are neglected, the series of corrections can be summed up to
infinity, giving the formula for $D(x)$ in a closed form, Eq. (\ref{4.29}),
described in Fig. 1. Then the equation describes stationary flow in a
quadratic potential. The theory works while $4mU''(x)<\gamma^2$, until the
averaged trajectory of a single particle is not oscillatory in the
potential wells along the 1D channel. The mapping in the oscillatory
regime requires the next analysis, which will be done in the future.

Technically, the paper demonstrates that the projection technique developed
for mapping of diffusion in 2D (3D) channels with varying cross section
\cite{map,exact,eff}, can be adapted for the dimensional reduction of a
process described by an evolution equation of a different type than the
diffusion or Smoluchowski equation. The method has been modified
significantly; $m/\gamma$ had to be confirmed as the small parameter
controlling the expansion of the corrections $\hat Z_k$, as well as
the operators of the backward mapping, $\hat\omega$ and $\hat\eta$.
In contrast to diffusion, the flux $j(x,t)$ is handled here as a 
quantity independent of the density $p(x,t)$. Thus the recurrence
procedure, calculating expansions of the correction operators $\hat Z_k$
and the operators $\hat\omega$, $\hat\eta$, is in principle the result
of combination of three equations, Eqs. (\ref{1.2}), (\ref{2.3}) and also
(\ref{2.5}), with the relation of the backward mapping, Eq. (\ref{2.9}).

It is worthwhile to notice that including the mass dependent corrections
to the Smoluchowski equation results in the equations (\ref{4.9}) or
(\ref{4.24}), which are of the same form as the comparable equations
obtained from the mapping of diffusion in channels with varying cross section.
On the other hand, the effective coefficient $D(x)$ (\ref{4.29}) has a
different symmetry than the similar formulas extending the Fick-Jacobs
equation \cite{eff,RR} for confined diffusion. Also $D(x)$ can be
greater than 1 here (see Fig. 1); i.e. the quasi stationary flux is
accelerated when passing through a shallow potential well, depending on
the nonzero mass of the particles. These interesting properties could be
observed in simulations similar to that verifying $D(x)$ in the extended
Fick-Jacobs equation \cite{Bnum,Dag}. The effects of slipping of the
particles diffusing under a nonconstant force $F(x)$ due to their inertia,
as described in Sect. IV, might also influence the interesting phenomena
in the micro and nano world, such as Brownian pumps \cite{BQAi1,BQAi2},
rectification of the flux in quasi 1D structures \cite{March}, 
stochastic resonance \cite{Bur1,Gamma,Bur2,GMSN}, or the negative mobility
\cite{HMSS}. Study of such applications of the theory presented is
expected in the future.

\section{Acknowledgments}
Support from VEGA grant No. 2/0049/12 and CE SAS QUTE project is
gratefully acknowledged. P.K. also thanks CIMS, New York University
for kind hospitality.

\renewcommand{\theequation}{A\arabic{equation}}
\setcounter{equation}{0}

\section{Appendix A: Exact solution}

The Green's function (\ref{3.2}) solving the FP equation with zero
potential, Eq. (\ref{3.1}), is calculated here. First we introduce
the scaled coordinates $\xi$, $u$, $\tau$ according to Eqs. (\ref{3.3})
and define the function $\Gamma(\xi,u,\tau;\xi',u',\tau')$,
\begin{equation}\label{a1}
G(x,v,t;x',v',t')=e^{-u^2/2}\Gamma(\xi,u,\tau;\xi',u',\tau')
e^{u'^2/2},
\end{equation}
satisfying the transformed equation (\ref{3.1}),
\begin{eqnarray}\label{a2}
\left[\partial_{\tau}+u\partial_{\xi}-\partial_u^2+u^2-1\right]
\Gamma(\xi,u,\tau;\xi',u',\tau')\hskip0.5in\cr
=\frac{\gamma\beta}{4}e^{(u^2-u'^2)/2}
\delta(\xi-\xi')\delta(u-u')\delta(\tau-\tau'),\ 
\end{eqnarray}
the exponential factor becomes unity due to $\delta(u-u')$.

After the Fourier transform in $\xi$ and $\tau$,
\begin{equation}\label{a3}
\Gamma(\xi,u,\tau;\xi',u',\tau')=\int\frac{dkd\nu}{4\pi^2}
e^{ik(\xi-\xi')-i\nu(\tau-\tau')}\Gamma_{k,\nu}(u;u'),
\end{equation}
and shifting the velocities by $ik/2$, $w=u+ik/2$ and $w'=u'+ik/2$,
the equation
\begin{equation}\label{a4}
\left[-i\nu-\partial_w^2+w^2-1+k^2/4\right]\Gamma_{k,\nu}(w;w')=
\frac{\gamma\beta}{4}\delta(u-u')
\end{equation}
becomes solvable if $\Gamma_{k,\nu}(w;w')$ is expressed in the basis 
set of the linear harmonic oscillator $\psi_n(w)$,
\begin{equation}\label{a5}
\Gamma_{k,\nu}(w;w')=\sum_{n=0}^{\infty}\Gamma_n(k,\nu)\psi_n(w)
\psi_n^*(w').
\end{equation}
The eigenfunctions $\psi_n(w)$ satisfy
\begin{equation}\label{a6}
\left(-\partial_w^2+w^2\right)\psi_n(w)=\lambda_n\psi_n(w)
=(2n+1)\psi_n(w)
\end{equation}
and we use the integral representation of the Hermite
polynomials $H_n(w)$ \cite{RG}
\begin{eqnarray}\label{a7}
\psi_n(w)&=&\frac{1}{\sqrt[4]{\pi}\sqrt{2^n n!}}H_n(w)e^{-w^2/2}\cr
&=&\sqrt{\frac{2^n}{n!\pi^{3/2}}}e^{-w^2/2}
\int_{-\infty}^{\infty}(w+ir)^ne^{-r^2}dr\hskip0.2in
\end{eqnarray}
in the next calculation.

Using the transformations above, we find
\begin{equation}\label{a8}
\Gamma_n(k,\nu)=\frac{\gamma\beta/4}{-i\nu+\lambda_n-1+k^2/4}.
\end{equation}
Applying it in the formulas (\ref{a5}) and (\ref{a3}), we integrate
the last one over $\nu$ in the complex plane,
\begin{eqnarray}\label{a9}
\Gamma(\xi,u,\tau;\xi',u',\tau')&&\hskip-0.2in=\frac{\Theta(\tau-\tau')}
{8\pi D_0}\int_{-\infty}^{\infty}dke^{ik(\xi-\xi')-k^2(\tau-\tau')/4}\cr
&&\times\sum_{n=0}^{\infty}e^{-2n(\tau-\tau')}\psi_n(w)\psi_n^*(w'),
\end{eqnarray}
$\Theta(x)$ denotes the Heaviside unit step function and
$D_0=1/\gamma\beta$ is the diffusion constant. Now the integral relation
(\ref{a7}) is used for $\psi_n(w)$ and $\psi_n^*(w')$ and the summation
over $n$ can be readily completed. Finally, the straightforward triple
integration over $k$, $r$, $r'$ is performed and using the transformation
(\ref{a1}) results in the formula (\ref{3.2}).

\renewcommand{\theequation}{B\arabic{equation}}
\setcounter{equation}{0}

\section{Appendix B: Quadratic approximation}

Derivation of the formula (\ref{4.29}) for the effective diffusion
coefficient $D(x)$ with all the derivatives higher than $U"(x)$ neglected
is presented here. This approximation corresponds to local replacing of
the potential by a parabola, $U(x)\simeq\kappa (x-x_0)^2/2+U_0$, where
$\kappa,\ x_0$ and $U_0$ are fitting parameters. 

First we simplify Eq. (\ref{4.27}). For quadratic potential,
the right hand side can be rewritten as
\begin{equation}\nonumber
e^{-\beta U(x)}\left(1+t_0\hat Z\right)^{-1}e^{\beta U(x)}=
\left(1+t_0 e^{-\beta U}\hat Z(x)e^{\beta U}\right)^{-1},
\end{equation}
$t_0\hat Z=\sum_{n=1}^{\infty}t_0\hat Z_n$; the difference contains only
the higher derivatives of $U(x)$, which are zero. Hence
\begin{equation}\label{b1}
D(x)/D_0=1+e^{-\beta U(x)}\sum_{n=1}^{\infty}t_0^n
\hat Z_ne^{\beta U(x)}.
\end{equation}

The formulas for $D(x)$ have been derived considering stationary
flow; $j(x,t)=j$ is constant. It simplifies the relation (\ref{4.7});
$\partial_tj=0$. Thus the right hand side represents stationary
flux, which can be directly compared with Eq. (\ref{4.8}), giving a
much simpler relation between $\hat Z_n$ and $\hat I_n$ than Eq.
(\ref{4.10}),
\begin{equation}\label{b2}
e^{-\beta U(x)}\hat Z_n(x)\partial_xe^{\beta U(x)}p(x)=
\partial_x\hat I_n(x)p(x)
\end{equation}
for any stationary solution $p(x)$. Calculation of the coefficients
of $D(x)$ according to Eq. (\ref{b1}) requires us to take $\partial_x
\exp[\beta U(x)]p(x)=\exp[\beta U(x)]$, hence finally
\begin{eqnarray}\label{b3}
e^{-\beta U(x)}\hat Z_ne^{\beta U(x)}&=&\frac{2}{\sqrt{\pi}}
\partial_x\int_{-\infty}^{\infty}u^2due^{-u^2}\hskip0.8in\cr
&&\times\hat\omega_n(x,u)e^{-\beta U(x)}\int dx e^{\beta U(x)}
\end{eqnarray}
after application of Eq. (\ref{4.6}).

Before writing the explicit formulas for $\hat\omega_n$ for the quadratic
potential, we define the polynomials
\begin{eqnarray}\label{b4}
P_n(u)&=&\sum_{k=0}^{n}\frac{(-1)^{n-k}2^{2k-n}}{(2k)!(n-k)!}u^{2k},\cr
Q_n(u)&=&\sum_{k=0}^{n}\frac{(-1)^{n-k}2^{2k-n}}{(2k+1)!(n-k)!}u^{2k},
\end{eqnarray}
$n=1,2, ...$, coming from the expansions of $\hat\omega_n$ and
$\hat\eta_n$ for zero potential in $t_0$, Eqs. (\ref{3.15}).
The first few polynomials are visible in the round brackets of Eq.
(\ref{3.19}). One can check by direct integration that
\begin{eqnarray}\label{b5}
\int_{-\infty}^{\infty}Q_n(u)u^2e^{-u^2}du&=&\frac{\sqrt{\pi}}
{2^{n+1}}\sum_{k=0}^n\frac{(-1)^{k-n}}{k!(n-k)!}=0,\cr
\int_{-\infty}^{\infty}P_n(u)u^2e^{-u^2}du&=&
\frac{\sqrt{\pi}}{2}\delta_{n,1},
\end{eqnarray}
corresponding to the normalization of $\hat\eta_n$, Eq. (\ref{4.5}),
and the relations (\ref{3.16}), (\ref{3.17}), proving no correction
to the Smoluchowski equation in the case $U(x)=0$.

The operators $\hat\omega_n$ and $\hat\eta_n$ for the quadratic
potential have the form
\begin{eqnarray}\label{b6}
\hat\omega_n&=&e^{-\beta U(x)}\sum_{k=1}^nc_{n,k}P_k(u)
\left(\beta U''\right)^{n-k}\partial_x^{2k}e^{\beta U(x)},\hskip0.3in
\cr \hat\eta_n&=&e^{-\beta U(x)}\sum_{k=1}^nc_{n,k}Q_k(u)
\left(\beta U''\right)^{n-k}\partial_x^{2k}e^{\beta U(x)},
\end{eqnarray}
with the coefficients
\begin{equation}\label{b7}
c_{n,k}=D_0^n\frac{2k\ (2n-1)!}{(n-k)!(n+k)!}.
\end{equation}
Due to the integrals, Eq. (\ref{b5}), only the first terms with $P_1(u)$
in Eq. (\ref{b6}) contribute to the expansion of $D(x)$, Eq. (\ref{b3}).
Then the functions become
\begin{eqnarray}\label{b8}
e^{-\beta U(x)}\hat Z_ne^{\beta U(x)}&=&c_{n,1}\partial_xe^{-\beta U(x)}
\left(\beta U''\right)^{n-1}\partial_xe^{\beta U(x)}\cr
&=&\frac{2\ (2n-1)!}{(n-1)!(n+1)!}\left(D_0\beta U''\right)^n,
\end{eqnarray}
taking $U^{(3)}(x)=0$ into account. Applied in Eq. (\ref{b1}) it results
in the expansion of $D(x)$, Eq. (\ref{4.29}).

Finally, one has to verify that the formulas (\ref{b6}) satisfy the
recurrence relations (\ref{4.13}) and (\ref{4.17}), acting on the function
$p(x)=\exp[-\beta U(x)]\int dx\exp[\beta U(x)]$. Although the equations
simplify notably due to neglecting the derivatives higher than $U''(x)$,
we omit the details of this tedious but straightforward calculation.

\end{document}